\documentclass{PoS}
\usepackage{amsmath,bm,amssymb,color,mathtools}
\usepackage{xspace}
\usepackage{graphicx}
\usepackage{multirow}
\usepackage{cite}

\DeclareRobustCommand{\NNLOJET}{{\normalfont\textsc{NNLOjet}}\phantom{j}}
 
\DeclareRobustCommand{\rT}{\mathrm{T}}  
\DeclareRobustCommand{\GeV}{\mathrm{GeV}\xspace}

\DeclareRobustCommand{\PZ}{{\mathrm{Z}}\xspace}
\DeclareRobustCommand{\PW}{{\mathrm{W}}\xspace}

\DeclareRobustCommand{\ptz}{{p^{\PZ}_{\rT}}\xspace}
\DeclareRobustCommand{\Pl}{{{\ell}}\xspace}

\DeclareRobustCommand{\mll}{{m_{\Pl\Pl}}\xspace}
\DeclareRobustCommand{\order}[1]{{{\mathcal{O}\!\left(#1\right)}}\xspace}
\DeclareRobustCommand{\alphas}{{\alpha_{\mathrm{s}}}\xspace}
\DeclareRobustCommand{\phistar}{{\phi^{*}_{\eta}}\xspace}

\def\ys{$|y^*|$\xspace}

\def\Q2{\left(Q^{2}\right)}

\def\l({\left(}
\def\r){\right)}

\def\ba{\begin{eqnarray}}
\def\ea{\end{eqnarray}}

\def\l{(}
\def\r{)}

\title{
\vspace*{-1.9cm}
\begin{minipage}{\textwidth}
{\normalfont\small \hspace{\fill} CFTP/18-001 }
\end{minipage}\\[20pt]
Jet cross sections and transverse momentum distributions with NNLOJET}

\ShortTitle{Jet cross sections with NNLOJET}

\author{\speaker{T.~Gehrmann}, X.~Chen\\
Department of Physics, University of Z\"urich, CH-8057 Z\"urich, Switzerland}

\author{J.~Cruz-Martinez, J.R.~Currie, E.W.N.~Glover, T.A.~Morgan, J.~Niehues, D.M.~Walker\\
Institute for Particle Physics Phenomenology, Department of Physics, University of Durham, Durham, DH1 3LE, UK}

\author{R.~Gauld, A.~Gehrmann-De Ridder\\
Institute for Theoretical Physics, ETH, CH-8093 Z\"urich, Switzerland} 

\author{A.~Huss\\
CERN Theory Division, CH-1211 Geneve 23, Switzerland} 

\author{J.~Pires\\
Centro de Fisica Teorica de Particulas - CFTP, 
Instituto Superior Tecnico IST,
Universidade de Lisboa, Av. Rovisco Pais,
P-1049-001 Lisboa, Portugal}

\abstract{This talk discusses recent results for next-to-next-to-leading order (NNLO) QCD corrections
to jet cross sections and transverse momentum distributions. The results are obtained in the \NNLOJET code 
framework, which provides an implementation of the antenna subtraction method for the handling 
of infrared singular contributions at NNLO. We briefly describe the \NNLOJET implementation, 
with particular emphasis on the construction of the real radiation phase space, which is tailored to ensure 
stability in all infrared sensitive regions. }

\FullConference{13th International Symposium on Radiative Corrections\\
		24-29 September, 2017\\
		St. Gilgen, Austria}

\begin{document}

\section{Introduction}
Low multiplicity final states containing jets and/or gauge bosons are produced copiously at hadron colliders. Production 
cross sections and kinematical distributions can be measured to high accuracy, thereby enabling precision studies 
of Standard Model parameters. These studies rely on a detailed comparison of the  experimental measurements 
with theory predictions, that are therefore required to match the accuracy of the experimental data. For many benchmark 
observables, this implies going beyond the common standard of next-to-leading order (NLO) in  QCD and the electroweak 
theory. Important progress has been made in the past years in the calculation of next-to-next-to-leading order (NNLO) 
QCD corrections to collider observables~\cite{twogamma,vh,vgamma,czakon1,czakon2,hjet,ourhj,wjet,wpt,ourzj,zpt,zphi,zang,zjet,mcfmgam,zz,ww,zw,abelof,2jnew,our2j,disprl,dis2j}. In this talk, we report some results obtained 
using the \NNLOJET framework, which is an implementation of the antenna subtraction method~\cite{ourant,currie} for 
NNLO QCD calculations.

\section{The NNLOJET framework}

At NNLO in QCD, collider observables receive three generic types of corrections to the basic 
Born-level process: double real (RR) corrections, real-virtual
(RV) corrections and double virtual (VV) corrections.
They are individually infrared divergent, and only their sum becomes finite. 
To obtain predictions that are fully differential in the final state kinematics, and that take proper account 
 of the experimental jet reconstruction algorithm, the three contributions need to be processed separately. Several techniques 
 for the extraction and recombination of infrared divergent contributions have been derived and  
 implemented in actual calculations~\cite{ourant,currie,secdec,qtsub,stripper,njettiness,trocsanyi}. 
 \NNLOJET uses the antenna 
 subtraction method, which constructs infrared subtraction terms for the RR, RV and VV contributions from so-called 
 antenna functions~\cite{ourant,currie,nloant,daleo,monni} that describe the unresolved parton radiation off a pair of hard radiator partons. 

The \NNLOJET code is a parton-level event generator that provides the framework for the implementation 
of jet production processes to NNLO accuracy, using the antenna subtraction method. It contains 
the event generator infrastructure (Monte Carlo phase-space integration, event handling and analysis routines) and 
provides the unintegrated and integrated antenna functions and the phase-space mappings for all kinematical 
situations. The implementation of processes in the \NNLOJET framework requires the availability of the matrix 
elements for all RR, RV, and VV processes, as well as 
 the construction of the antenna subtraction terms. \NNLOJET provides 
testing routines to verify the point-wise convergence of the subtraction, 
as documented for example in Ref.~\cite{joao}. Processes included in \NNLOJET  up to now 
are $Z$ and $Z+j$ production~\cite{ourzj,zpt,zphi,zang}, $W$ and $W+j$ production~\cite{wpt}, 
$H$ and $H+j$ production~\cite{ourhj}, 
 di-jet production in hadron-hadron collisions~\cite{2jnew,our2j} and in lepton-hadron collisions~\cite{disprl,dis2j}, as
 well as three-jet production in electron-position annihilation~\cite{our3jnew}.  
We discuss several recent phenomenological applications in Section~\ref{sec:appl} below. 

\NNLOJET supports parallel computing infrastructures through the OpenMP interface, which is 
especially crucial for the Monte Carlo warmup (adaptation of integration grids). Typical runtimes for 
NNLO predictions of kinematical distributions in $2\to 2$ processes range between 80'000 and 250'000 core-hours, with most 
of the computation time spent on the double real radiation processes.  

\section{Phase space generation at NNLO}

In using the antenna subtraction method to construct subtraction terms 
for higher order calculations, one encounters the problem of angular correlations 
in the collinear 
splitting of a gluon into massless partons. These angular correlations
introduce non-factorizing terms which correlate the hard reduced matrix 
element with the splitting functions. They vanish when the 
azimuthal variable of the collinear system (with respect to the collinear 
axis, defined by the collinear momentum and a light-like recoil momentum) 
is integrated out. The same cancellation can also be accomplished~\cite{weinzierl}
 by combining two collinear configurations related 
by an azimuthal rotation of $\pi/2$. To specify the frame in which the rotation is performed, 
two light-like directions must be specified: one is given by the collinear momentum of the pair, while the other 
can be chosen arbitrarily. 

The \NNLOJET phase space generators for different processes produce pairs of phase space 
points related by angular rotation around a specified axis. In the following, we 
describe the phase space implementation using vector-boson-plus-jet production as 
an example process. 
We consider the basic kinematical situation 
\begin{displaymath}
p_a + p_b \to p_1 + p_2 +p_3 (+p_4) (+p_5)
\end{displaymath}
where the outgoing  momenta $p_1$ and $p_2$ represent the decay leptons that are not 
involved in any unresolved limit. 

To denote the angular rotations, we introduce a shorthand notation. 
At NLO, we can have one collinear splitting, and thus 
one collinear pair to be rotated. The notation
\begin{list}{}{\labelwidth 1.5cm \labelsep 0.3cm \leftmargin 1.8cm}
\item[{$(i,j;k)$}] represents the rotation of the momentum pair  
$p_i$, $p_j$ around the axis defined by $p_k$. The corresponding phase
space contains $s_{ij}$ as one of its basic variables. 
\end{list}
The result is two complete sets of final state momenta: unrotated and 
rotated. 

At  NNLO, two pairs of partons can become 
collinear, or three partons can become simultaneously collinear. We 
distinguish two cases:
\begin{list}{}{\labelwidth 3cm \labelsep 0.3cm \leftmargin 3cm}
\item[{$(i;j,k;l)$}] is the rotation appropriate to triple collinear splitting. 
The momenta $p_j$ and $p_k$ are first rotated around the $p_l$ axis, and the 
resulting system of $p_i$, $p_j$, $p_k$ is then also rotated around the $p_l$ 
axis. The phase space is constructed from a sequential splitting 
$(ijk)\to i+(jk) \to i+j+k$ with $s_{ijk}$ and $s_{jk}$ as the basic 
phase space variables.
\item[{$(i,k;l,m;n)$}] is the rotation appropriate to double 
single-collinear splitting. The momenta $p_i$, $(p_l+p_m)$ are 
rotated around the $p_k$ axis, while the momenta $p_l$, $p_m$ are rotated 
around the $p_n$ axis. The phase space is again constructed from a sequential splitting 
$(ilm)\to i+(lm)\to i+l+m$.
\end{list}
In both cases, the output consists of four complete sets of final state momenta:
the original momenta, only the first pair rotated, only the second 
pair rotated, and both pairs rotated.

The implemented phase space parametrizations are thus optimized to 
account for the angular rotations in a specific single or double 
unresolved limit. They are not appropriate for the fully inclusive coverage of 
the phase space. Instead, we introduce phase space wedges, which are 
specified through a set of selection criteria, such that the full 
final state phase space of a given multiplicity 
can be obtained by summing over all possible wedges. Each wedge only
contains those unresolved limits to which the angular rotation is 
appropriate.
The weight from the phase space generator is the same for the original and rotated momentum set, 
while the matrix element and subtraction terms are evaluated separately for both sets. 
  The final weight 
is obtained by summing over all events (wedges and their rotations), 
and dividing by the number (2 or 4) of angular rotated 
partner events. 

The wedges are defined by imposing constraints on the invariants, which for this purpose 
are taken to be the $s_{ij} = 2 p_i\cdot p_j$ with $\{ p_i,p_j\} \in \{p_a,p_b,p_3,p_4,p_5\}$.
In the list of all $s_{ij}$, we denote the two smallest invariants by $s_1$ and $s_2$.

\subsection{Three-particle phase space}
The three-particle final state is free of singularities, as ensured
by the final state 
selection cuts, requiring the vector boson and a jet at finite transverse momentum.

\subsection{Four-particle phase space}
The four-particle contribution to vector-boson-plus-jet production contains simple collinear initial or 
final state radiation. To properly account for the corresponding angular terms in some of the 
limits, we consider the rotation $(3,4;a)$. 
This rotation takes proper account of the angular terms in the collinear 
limits $(3\parallel 4)$, $(3\parallel a)$ and $(4\parallel a)$. The phase space 
 routine  therefore fails to take proper account of the angular terms in 
the initial state collinear limits involving $p_b$. Therefore, we restrict it to the following phase space wedge 
(excluding $s_{34}$ from the ordered list of invariants):
\begin{eqnarray*}
&&s_1=s_{a3} \; \mbox{or}\;  s_1 = s_{a4} 
\end{eqnarray*}
which singles out the three above-mentioned collinear limits, and can therefore also be 
identified by  the notation $(3,4;a)$. The full phase space is 
then obtained by summing over two wedges:
$(3,4;a)$, $(3,4;b)$, the angular average is obtained by 
averaging over the angular partner events in each wedge.

\subsection{Five-particle phase space}
The five-particle contribution to vector-boson-plus-jet production can contain triple-collinear and double 
single-collinear configurations. We decompose the full five-particle phase space into 
6 triple-collinear and 6 double single-collinear wedges. The full phase space is
recovered by summing over all wedges. The phase space integration is structured such that 
two separate integrals (1.\ sum of triple-collinear wedges and 2.\ sum of double single-collinear wedges) 
must be evaluated to obtain the full five-particle phase space. 

For the
angular terms in triple-collinear configurations, we consider the rotation $(3;4,5;a)$
This rotation takes proper account of the angular terms in the limits $(3\parallel 4 \parallel 5)$,
$(a\parallel 4\parallel 5)$ and ($4\parallel 5,\, 3$ soft). To ensure that 
only these limits are covered by the phase space generator, its application is restricted to a particular wedge, 
which is defined by 
\begin{eqnarray*}
&& \left( \{s_1,s_2 \}  \in \{s_{45},s_{a4},s_{a5}\}\right)\; \mbox{or}\\
&& \left(s_1=s_{45}\; \mbox{and}\; s_2 = \mbox{min}(s_{34},s_{35}) \; \mbox{and}\; s_{a4}+s_{a5} < s_{b4} + s_{b5} \right) \; .
\end{eqnarray*}

All six triple-collinear wedges are then obtained from $(i;j,k;a)$ by permuting $i$ over the 
three partonic final state momenta, and interchanging $(a,b)$.
Full 
phase space coverage for the triple-collinear 
wedges is obtained by summing the six wedges, and the angular average is obtained by 
averaging with the angular partner events. 

To properly account for the angular terms in double single-collinear configurations, we consider the 
rotation $(3;a;4,5;b)$. This rotation takes proper account of the angular terms in 
$(3\parallel a;4\parallel 5)$, $((4\mbox{ or } 5)\parallel b;3\parallel a)$. To ensure that only these 
limits  are covered by the phase space generator, its application is restricted to a particular wedge, 
which is defined by 
\begin{eqnarray*}
&& \left( s_1 = s_{a3}\; \mbox{and}\; s_2 = \mbox{min}(s_{b3},s_{b4},s_{b5}\right)\; \mbox{or}\\
&& \left( \{s_1,s_2 \}  \in \{s_{45},s_{a3}\}\right)\; .
\end{eqnarray*}

All 6 double single-collinear wedges are then obtained from $(i;a;k,l;b)$ by permuting 
$i$ over the three partonic final state momenta and interchanging $(a,b)$. 
 Full 
phase space coverage for the double single-collinear 
wedges is obtained by summing the six wedges, the angular average is obtained by 
averaging with the angular partner events. 

The full phase space is obtained by summing the integration results from the 
sum of the triple-collinear wedges and the sum of the double single-collinear wedges.

\section{Recent applications}
\label{sec:appl}

\subsection{Transverse momentum distributions in $Z$ boson production}
The production of $\PZ$-bosons which subsequently decay into a pair of leptons is a Standard Model benchmark 
process at hadron colliders. It occurs with a large rate and can be measured with small experimental uncertainties due to its 
clean final state signature. It has been studied extensively at the LHC by the ATLAS~\cite{ptzATLAS7TeV,ptzATLAS} and CMS~\cite{ptzCMS7TeV,ptzCMS}  experiments. 
A key observable in these measurements is the transverse momentum distribution of the $\PZ$-boson, which 
provides direct access  to the gluon distribution in the proton. 
The transverse momentum of the $\PZ$-boson is due to the emission of QCD radiation from the initial state partons. As a consequence, fixed order predictions at $\order{\alphas^2}$ in perturbative QCD, which 
are NNLO-accurate for the inclusive $\PZ$-boson production cross section, 
correspond only to NLO-accurate predictions for the transverse momentum distributions. 
The perturbative description of the 
transverse momentum distribution of the $\PZ$-boson is therefore most closely related to 
$\PZ$+jet production, with the jet reconstruction replaced by a transverse momentum cut on the $\PZ$-boson.

Using \NNLOJET, we computed the transverse momentum distributions for $\PZ$-boson production~\cite{zpt,zphi}, 
and most recently also for $\PW$-boson production~\cite{wpt}. 
In the small $\ptz$ region, the precision of direct measurements of the $\ptz$ spectrum using the standard $\ptz$ variable is limited by the experimental resolution on $\ptz$ itself, and in particular on the resolution of the magnitude of the transverse momenta of the individual leptons entering $\ptz$.
To probe the low $\ptz$ domain of $\PZ/\gamma^*$ production an alternative angular variable,
$\phistar$, has been proposed~\cite{banfidefphistar}. This 
variable is reconstructed entirely from the lepton directions (without using the lepton energies)
and therefore minimises the impact of these experimental uncertainties. 
At low $\ptz$, one finds the approximate relation:
 \begin{equation}
  2 \phistar \approx \ptz/\mll. 
  \label{eq:phiapprox}
\end{equation}

\begin{figure}
  \centering
  \includegraphics[width=.78\linewidth]{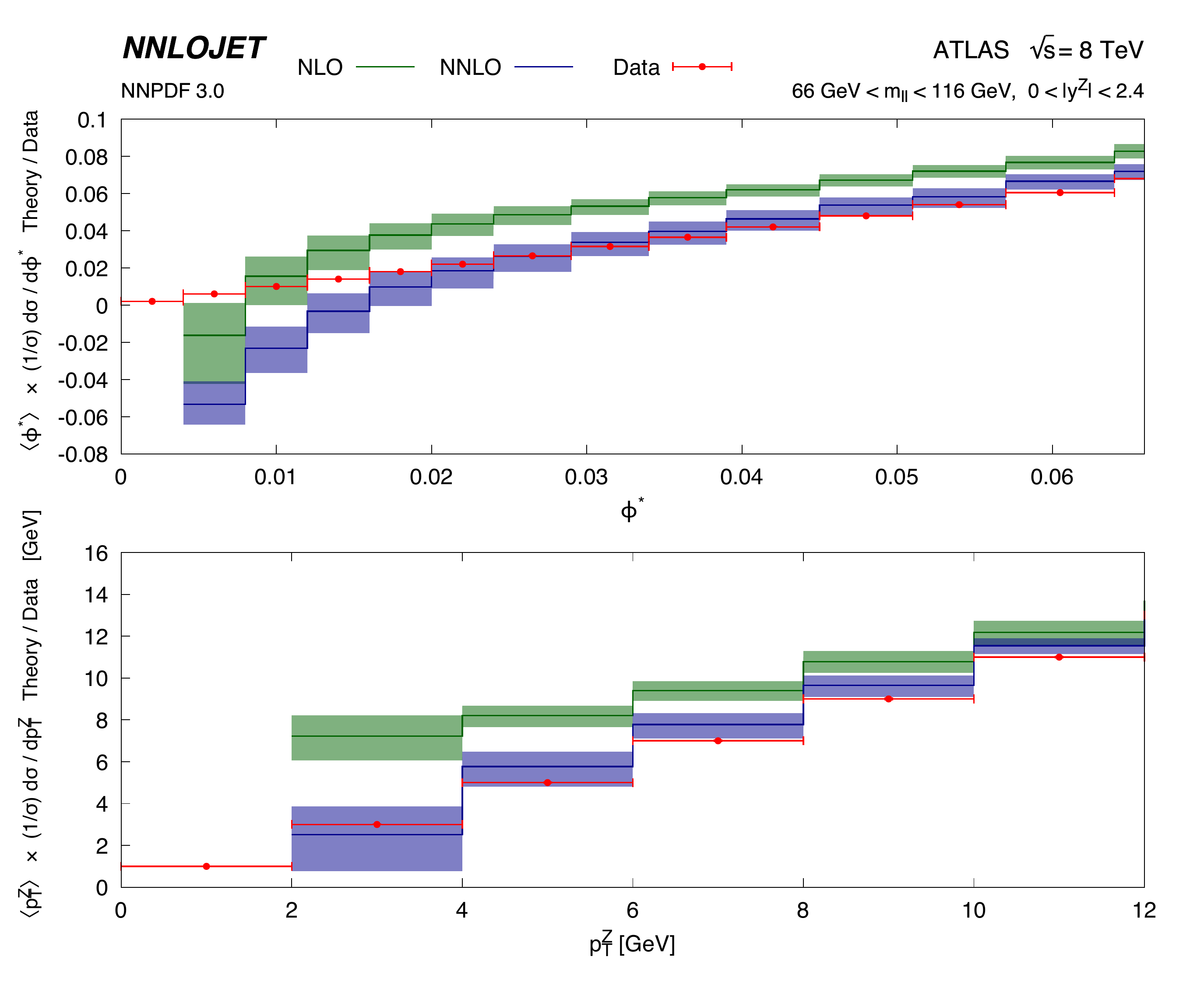}
  \caption{
 The $\phistar$ and $\ptz$ distributions for the on-resonance mass bin $66~\GeV<\mll<116~\GeV$.
The distribution is normalised to the experimental ATLAS data~\protect\cite{ptzATLAS}.  The green bands denote the NLO prediction with scale uncertainty and the blue bands show the NNLO prediction with scale uncertainty.    Figure from 
\protect\cite{zphi}.
  }
  \label{fig:lowphi_star_on_res}
\end{figure}
The  
distributions in $\phistar$ and $\ptz$ are therefore closely related in the infrared region. In particular, 
one should expect the onset of 
large logarithmic corrections (and consequently the breakdown of the fixed order NLO and NNLO predictions) to 
occur at values of $\phistar$ and $\ptz$ related through the above equation. 
To illustrate this,  Fig.~\ref{fig:lowphi_star_on_res} superimposes the infrared regions of these distributions. 
The $\ptz$ range is fixed to $[0, 12]~\GeV$, while the 
$\phistar$ range is chosen according to Eq.~\eqref{eq:phiapprox}. 
The first bins contain the zero value and are not accessible by a fixed-order calculation of the $\ptz$ or $\phistar$ distributions, which diverge there. 

We observe the substantially higher experimental resolution in $\phistar$. This reflects the much better experimental resolution of the low $\ptz$ region afforded by the $\phistar$ variable.
The NLO prediction fails to describe the data in the 
plotting range and only starts to agree with data for larger values of $\ptz$ and $\phistar$.
The NNLO description, on the other hand, remains reliable down to values of 
$\phistar\approx 0.02$. The 
precise point of deviation in the $\ptz$ distributions cannot be resolved due to the coarse binning. 
Nevertheless, the values of $\phistar$ and $\ptz$ where the fixed order predictions start to deviate from the data appear to be in line with the 
expectation from Eq.~\eqref{eq:phiapprox}. A description of the distributions in $\phistar$ and $\ptz$ 
over the full kinematical range will require  
the matching of the fixed order NNLO predictions onto resummation. 

Even more detailed information on the production dynamics of $\PZ$-bosons can be extracted from the angular coefficients 
that determine the full kinematical distribution of the decay leptons. Also for these angular coefficients, we observe that 
inclusion of NNLO QCD corrections results in a stabilization of the theoretical predictions and 
a considerably better description of the experimental data~\cite{zang}. Together with the 
transverse momentum distributions of $\PZ$ and $\PW$ bosons, these angular coefficients 
 play an outstanding role in the determination of the $\PW$-boson mass from lepton-plus-missing energy distributions. 
 With our \NNLOJET implementations~\cite{wpt,zpt,zphi,zang}, predictions for all these distributions
 can now be obtained to NNLO accuracy.

\subsection{Jet production in deep inelastic scattering}

Our understanding of the inner structure of the proton has been shaped through a long series of deep-inelastic lepton-nucleon
scattering (DIS)
experiments, which have established the partonic structure of the proton and provided precision measurements of 
parton distribution functions (PDFs). While the quark distributions can be probed directly in inclusive DIS, 
gluon-initiated processes enter only as higher order corrections. A direct determination of the gluon distribution
requires the selection of specific hadronic final states such as heavy quarks or jets. 

The DESY HERA electron-proton collider provided a large data set of hadronic final states in 
DIS at $\sqrt{s} = 319$~GeV. Jet final states have been measured to high precision over a large kinematical range 
 by the H1~\cite{h1highq2,h1lowq2} and 
ZEUS~\cite{zeus1}  experiments. 
The reconstruction of jets is performed in the Breit frame, defined by the 
direction of the virtual photon and incoming proton. Jet production in the Breit frame is induced both by  
quarks and gluons in the initial state, with gluons making up the dominant contribution almost throughout the 
entire kinematical range. 
Up to very recently, the theoretical description of jet production in DIS was limited to NLO in QCD. At this order, 
the residual theory uncertainty is typically larger than the experimental errors on the HERA data, thereby 
limiting the impact of the data in precision QCD studies.
\begin{figure}
\centering
  \includegraphics[width=0.9\linewidth]{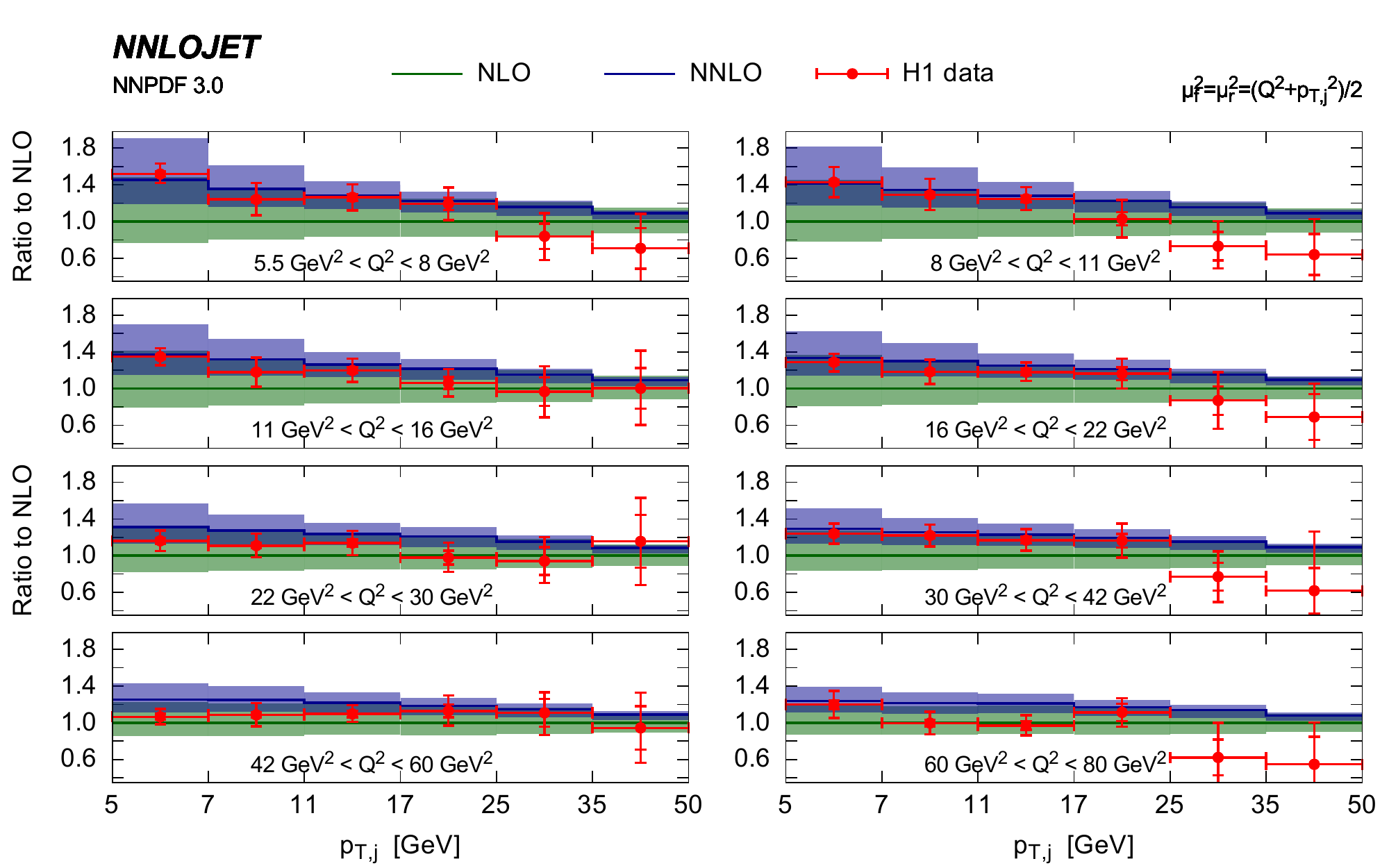}
    \caption{
Inclusive jet production cross section as a function of the jet transverse momentum $p_{T,B}$ in bins of $Q^2$, compared to H1 data~\protect\cite{h1lowq2}. Figure from~\protect\cite{dis2j}.} 
  \label{fig:h1inc}
\end{figure}

Using the \NNLOJET framework, we computed the NNLO corrections to jet production in DIS, both for di-jet 
final states~\cite{disprl} and for single inclusive jet production~\cite{dis2j}. 
Figure~\ref{fig:h1inc} compares our NNLO predictions to the H1 single jet inclusive data~\cite{h1lowq2}. 
We observe that the NNLO corrections are very substantial at low-$Q^2$ and
low-$p_T^B$, with an up to 60\% enhancement with respect to NLO. These large corrections are within the NLO 
uncertainty band (close to the upper edge), and 
lead to in a residual theory uncertainty of 20\% even at NNLO. Especially at low $Q^2$, the shape and normalisation of the theory prediction 
changes significantly going from NLO to NNLO, and results in a considerably improved theoretical description of the data.

Our newly derived NNLO predictions were used for a re-analysis of all H1 jet and di-jet data in view of 
an improved determination of the strong coupling constant and the gluon distribution~\cite{h1as}. This study 
is the first application of an \NNLOJET interface to the ApplFast-NNLO framework that recasts the NNLO predictions 
in terms of parton-level coefficient functions, thereby enabling fast multiple evaluations of the predictions for 
varying coupling constants, scales and parton distribution functions. 

Using parton distributions from the global NNPDF3.1 fit~\cite{nnpdf}, the re-analysis of the H1 jet data yields the determination of 
the strong coupling constant~\cite{h1as}:
$$\alpha_s(M_Z,\mbox{H1 jets, NNLO}) = 0.1157(20)_{{\rm exp}}(29)_{{\rm th}}.$$
Based on the same data set, a simultaneous determination of the strong coupling constant and the parton 
distribution functions was also performed~\cite{h1as}, resulting in a slightly lower value of $\alpha_s$ with quark and 
gluon distributions rising more steeply at low $x$ than observed in the global fits.

\subsection{Jet production at LHC}
Hadron-hadron collisions generally lead to final states containing jets. 
When at least two jets are produced, the two jets leading in transverse momentum, $p_{T}$, constitute a 
dijet system. The two jets in the final state allow for a full reconstruction of parton-level kinematics, thereby 
providing 
valuable information on important Standard Model parameters such as the strong coupling, $\alpha_{s}$, and
the PDFs. Di-jet production is being studied in detail by the LHC
 experiments~\cite{cmsdijet,atlasdijet}, both in view of searches for strongly interacting physics beyond the 
 Standard Model and for precision measurements. 

To fully exploit the experimental data, it is important to have a reliable and accurate theoretical prediction. 
Up to very recently, jet production at hadron colliders was known to 
NLO accuracy in perturbative QCD. 
Although the NLO corrections give an improvement on the LO prediction, there remains significant theoretical
uncertainty associated with the NLO calculation. It is well known that the choice of scales for renormalization, $\mu_{R}$,
and factorization, $\mu_{F}$, 
has a sizable impact on the predictions at NLO and, for this reason, the dijet data is regularly excluded from global PDF fits.
 \begin{figure}[t]
  \centering
    \includegraphics[width=0.45\textwidth]{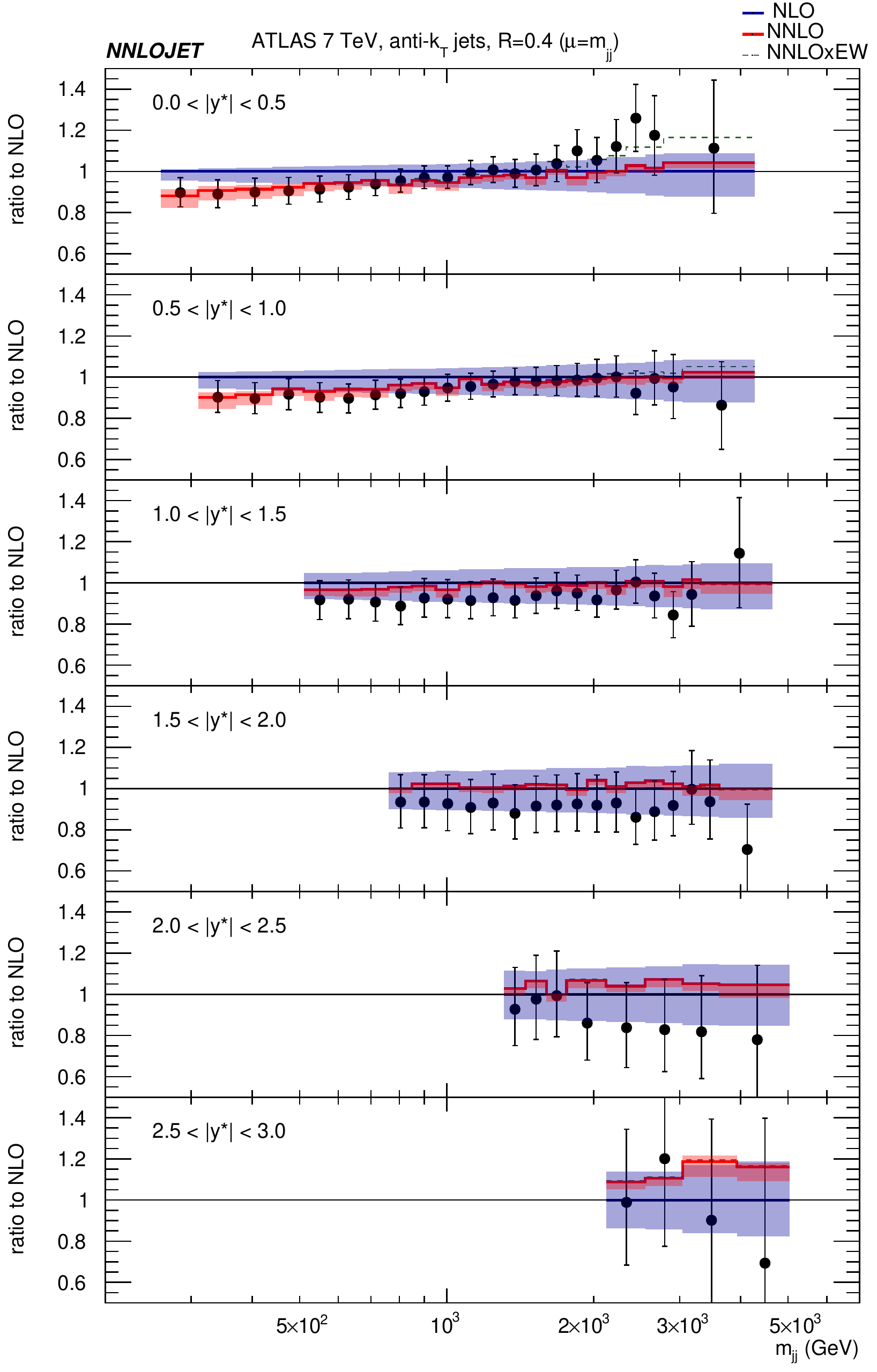}
  \caption{The NLO (blue) and NNLO (red) theory predictions and ATLAS data~\protect\cite{atlasdijet} 
  normalized to the NLO central value. The bands represent
  the variation of the theoretical scales in the numerator by factors of 0.5 and 2. Electroweak effects are implemented as a multiplicative factor and 
  shown separately as the green dashed line. Figure from~\protect\cite{our2j}.}
  \label{fig:rat2nlo}
\end{figure}

NNLO corrections to jet production at hadron colliders were first computed using \NNLOJET for 
single jet inclusive production~\cite{2jnew}, and most recently extended to 
 di-jet observables~\cite{our2j}. We observe that the inclusion of the NNLO corrections 
 results in a considerable reduction of the spread of the predictions obtained for different central scale choices.  
 Fig.~\ref{fig:rat2nlo} shows the 
 di-jet invariant mass distributions at 7~TeV in different bins in rapidity, normalized to the NLO prediction. 
Comparison with the ATLAS data~\cite{atlasdijet} shows good 
agreement with the NNLO
QCD prediction across the entire dynamical range in $m_{jj}$ and \ys and a significant improvement in the description of the data for low $m_{jj}$ and
\ys, where NLO does not adequately capture the shape nor the normalization. We include the electroweak effects as a multiplicative factor, as calculated
in~\cite{eweak}, and note that 
they improve the description of the data at central rapidity and high invariant masses  (\ys$<0.5$, $m_{jj}>2$~TeV).
It will now be very interesting to investigate the impact of these data on a global NNLO determination of parton distributions.

\section{Conclusions and Outlook}

In this talk, we highlighted several recent results for NNLO QCD corrections to 
jet observables and transverse momentum distributions that were obtained using the \NNLOJET framework. In general, 
it is observed that the NNLO predictions provide a much-improved description of the kinematical distributions observed in 
experimental measurements, along with a substantial reduction of the residual theoretical uncertainty. 

The calculation of NNLO QCD corrections is computationally expensive. To use these results as input to an 
experimental analysis,  new methods for their dissemination will have to be explored. The NNLO study of H1 
deep inelastic jet data, leading to a new 
extraction of the strong 
coupling constant and a study of parton distributions, is a first example of such an application, using grid tables to parametrize
the NNLO results in terms of coefficient functions.

Moreover, we note that NNLO QCD calculations are currently limited by the availability of two-loop matrix elements to 
$2\to 2$ processes. Further progress will require a higher degree of automation in the calculation of two-loop 
matrix elements at higher multiplicity. 

\section*{Acknowledgements}

This research was supported in part by the UK Science and Technology Facilities Council, by the Swiss National Science Foundation (SNF) under contracts 200020-175595, 200021-172478, and CRSII2-160814, by the Research Executive Agency (REA) of the European Union under the Grant Agreement PITN-GA-2012-316704 (``HiggsTools'') and the ERC Advanced Grant MC@NNLO (340983), and  by the Fundacao para a Ciencia e a Tecnologia (FCT-Portugal), project UID/FIS/00777/2013.

\end{document}